\documentstyle[12pt]{article}

\begin{document}

\author{G. G. Varzugin$^\ast$ , A. S. Chistyakov$^\dag$}
\title{CHARGED ROTATING BLACK HOLES IN
EQUILIBRIUM}
\date{}
\maketitle

\begin{center}$^\ast${\it Laboratory of Complex System Theory, Physics Institute of
St-Petersburg University,\\ 198504, St. Petersburg, Peterhof,
Ulyanovskaya 1, Russia. E-mail
varzugin@paloma.spbu.ru}\end{center}
\begin{center}$^\dag${\it Department of Physics, St. Petersburg University, St. Petersburg, Russia.}\end{center}
\vskip1cm \abstract{Axially symmetric, stationary solutions of the
Einstein-Maxwell equations with disconnected event horizon are
studied by developing a method of explicit integration of the
corresponding boundary-value problem. This problem is reduced to
non-leaner system of algebraic equations which gives relations
between the masses, the angular momenta, the angular velocities,
the charges, the distance parameters, the values of the
electromagnetic field potential at the horizon and at the symmetry
axis. A found solution of this system for the case of two charged
non-rotating black holes shows that in general the total mass
depends on the distance between black holes. Two-Killing reduction
procedure of the Einstein-Maxwell equations is also discussed.}
\pagebreak

\section{Introduction}

In this paper we study axially symmetric, stationary solutions of
the Einstein- Maxwell equations having disconnected event horizon
(N black holes). The solutions of this kind have been known since
the early days of the general relativity Ref.~\cite{weyl} and,
from our point of view, are among the most interesting exact
solutions of the Einstein equations.  A great merit of these
solutions is that they have a clear mechanical interpretation. It
allows us to hope that the solutions of this class will help us to
understand N-body problem which, we think, is one of the most
fundamental and difficult problem of the classical general
relativity. We begin from description of the main physical
properties of these solutions.

In general the solutions of this family are parameterized by the
mass, the charge, the angular momentum of each of the black hole
and by the distances between them. Conical singularities at the
non-extreme symmetry axis components prevent the fall of the black
holes on each other. Interpreting these symmetry axis components
as 'matter strings' we get a natural definition of the interaction
force between the black holes, namely, the tension of 'string'. It
is exactly equal to the deficit angle of conical singularity
Ref.~\cite{weyl}(see also Ref.~\cite{weinstein90}). The
interaction force of two non-charged black holes tends to the
Newtonian limit as the distance parameter becomes large
Refs.~\cite{weyl,varzugin98} and goes to infinity as the distance
parameter approaches the value for which two components of the
horizon intersect Ref.~\cite{weinstein94}. The generalization of
these results for the case of charged black holes hasn't been done
yet.

It is worth mentioning that for so-called extreme black holes (the
horizon is contracted to points) the interaction force can be
equal to zero Ref.~\cite{hartle}. Unfortunately, this is the only
known case when the balance between the gravity attraction force
and the electromagnetic repulsion force takes place.

The total mass of N Schwarzschild black holes in equilibrium
doesn't depend on the distances. However, it is a function of the
distance parameters for the rotating black holes that is
stipulated by the spin-spin interaction of the black holes. The
charge 'densities' of the black holes must also effect on the
total mass since some part of the energy is needed for the charge
redistribution in the final black hole. In the case of the static
model we find that the mass of two charged black holes depends on
the distance parameter when their charge 'densities' are
different. To determine the masses of the black holes one must
solve very complex constraint equations for the angular momenta
and the angular velocities of the black holes
Ref.~\cite{varzugin97}. A partial solution of these constraints
was found in Ref.~\cite{varzugin98}. We obtain the equations for
the physical parameters of the charged black holes in this paper.

Despite of the presence of the conical singularity the gravity
action on the stationary N black holes solutions remains finite.
It allows us to construct the thermodynamics for the system of N
black holes. The thermodynamic properties of this system are known
only for non-rotating black holes Ref.~\cite{costa}. The surface
area of the horizon of two Schwarzschild black holes increases
when the distance parameter decreases.

All stationary, axially symmetric solutions with disconnected
event horizon satisfy a certain boundary-value problem for the
Einstein-Maxwell equations. The boundary conditions of this
problem are the regularity conditions for the symmetry axis and
for the event horizon Ref.~\cite{carter}. The condition of
asymptotic flatness is a part of the event horizon definition. The
uniqueness and existence theorem for this problem was proven in
Refs.~\cite{weinstein90,weinstein94,weinstein92,litian} for
non-charged black holes, and in Ref.~\cite{weinstein96} for
charged ones.

The Einstein-Maxwell equations with two commuting symmetries
belong to a wide class of completely integrable equations in the
sense: they can be written as compatibility conditions for some
auxiliary linear system of equations
Refs~\cite{alekseev1,xan,negkra83,egk84}. Using this fact one can
derive 2N-soliton solution. The 2N-soliton solutions do not
necessarily include the family of the solutions studied in this
paper. However, we have showed Ref.~\cite{varzugin97} that N black
holes (non-charged) family is a subclass of the 2N-soliton family
of Belinskii-Zakharov Ref.~\cite{belzhah79}.

The 2N-soliton solutions of the Einstein-Maxwell equations can be
constructed by a slight modification of the Belinskii-Zakharov
method Ref.~\cite{egk84}. Different approaches were applied in
Refs.~\cite{alekseev1,negkra83,ernst79}. The 2N-soliton family of
Refs.~\cite{alekseev1,negkra83} can not contain the black holes
solutions (likely, the family of Ref.~\cite{ernst79} either). It
was extended in Ref.~\cite{manko95} with the use of the technique
suggested in Ref.~\cite{sib84}. A key problem with the 2N-soliton
solution is that this solution has extra parameters which must be
excluded from it to get N black holes solution. The only way to do
this is to satisfy the regularity conditions of the horizon and
the symmetry axis by imposing some constraints on the parameters.
In this paper we use these conditions at the beginning.

In the end of this section we wish to emphasize that physical
solutions for two black holes in equilibrium are those the
interaction force of which is equal to zero.

In the next section we state the boundary-value problem, which
defines N charged rotating black holes in equilibrium strictly.

\section{Boundary-value problem}

In this section we formulate the boundary-value problem for the
Einstein-Maxwell equations, which describes the all possible
axially symmetric, stationary solutions having disconnected event
horizon. Space-time manifold is said to be stationary and axial
symmetric if there are two commuting Killing fields of which one
is timelike, $k^\mu$, and the other is spacelike, $m^\mu$. The
vector fields $k^\mu$ and $m^\mu$ are generators of the isometry
groups which are isomorphic $R$ and $SO(2)$ respectively. We
denote these fields $K_0 = k$ and $K_1 = m$ either. By definition,
the vector fields $K_A$ satisfy the Killing equation and commute,
\begin{equation}\label{kiling}
\nabla_{\mu}K_{A\nu}+\nabla_{\nu}K_{A\mu}=0,\;\;\;[K_A,K_B]=0.
\end{equation}
Indexes denoted by the Latin capitals run 0 and 1.

It is more convenient to state the conditions for the self-dual
electromagnetic field $F^\dagger$, $$ F^\dagger=F+i\ast F,\;\;\;
\ast F_{\mu\nu}=\frac{1}{2}e_{\mu\nu\alpha\beta}F^{\alpha\beta},
$$ than for the electromagnetic field $F$. Here,
$e_{\mu\nu\alpha\beta}$ is the alternating tensor and $\ast$ is
the Hodge operator. We assume that $F^\dagger$ is invariant under
the action of the isometry groups, viz
\begin{equation}
{\cal L}_{K_A} F^\dagger=0,\label{groupinvar}
\end{equation}
where ${\cal L}_{K_A}$ is the Lie derivative in the direction
$K_A$, and the field circularity condition holds,
\begin{equation}
F^\dagger_{\mu\nu}K_A^\mu K_B^\nu=0.\label{cyk}
\end{equation}

Using $F^\dagger$ we can write the Einstein-Maxwell equations in
the form
\begin{equation}
R^\mu_{\enskip\nu}=8\pi T^\mu_{\enskip\nu},\;\;\;
T^\mu_{\enskip\nu}=\frac{1}{8\pi}F^\dagger_{\nu\alpha}\bar F^{\dagger\mu\alpha},
\;\;\;\nabla_{[\mu}F^\dagger_{\nu\alpha]}=0.
\end{equation}
Let us introduce notations
\begin{equation}
g_{AB}=K_{A\mu}K_B^\mu,\;\;\;
\Phi_{\alpha A}=F^\dagger_{\alpha\beta}K^\beta_A
\label{opg}
\end{equation}
and a matrix-valued twist potential
\begin{equation}
\omega_{AB}^{\alpha}=e^{\alpha\mu\nu\beta}K_{A\mu}\nabla_{\nu}K_{B\beta}.
\label{twist}
\end{equation}
From the Maxwell equations and the condition~(\ref{groupinvar}) we
get that
\begin{equation}
\Phi_{\alpha A}=\nabla_\alpha\Phi_A.
\end{equation}
A simple corollary of the definitions~(\ref{opg}) and
(\ref{twist}) is
\begin{equation}
e_{\mu\nu\theta\beta}\omega_{AB}^\beta K_C^\theta K_D^\nu=
g_{AC}\nabla_{\mu}g_{BD}-g_{AD}\nabla_{\mu}g_{BC}.\label{omegaproj}
\end{equation}

Let us now recall the basic properties of the Killing fields  $$
\nabla_\mu\nabla_\nu K_{\alpha
A}=R_{\beta\mu\nu\alpha}K^\beta_A,\;\;\; \nabla^\mu\nabla_\alpha
K_{\mu A}=R_{\alpha\beta}K^\beta_A, $$ which are consequence of
Eq.~(\ref{kiling}) and the curvature tensor definition. Using them
it is easy to check that
\begin{equation}
\nabla_{[\alpha}\omega_{\beta]AB}+
e_{\alpha\beta\mu\nu}K^\mu_A R^{\nu\gamma}K_{\gamma B}=0.
\label{twistproperty1}
\end{equation}
Furthermore, from the Einstein-Maxwell equations we see that $$
e_{\alpha\beta\mu\nu}K^\mu_A R^{\nu\gamma}K_{\gamma B}=
e_{\alpha\beta\mu\nu}K^\mu_A F^{\dagger\nu\delta} \bar
F^\dagger_{\gamma\delta}K_B^\gamma=
-ie_{\alpha\beta\mu\nu}K^\mu_A\ast F^{\dagger\nu\delta} \bar
F^\dagger_{\gamma\delta}K_B^\gamma $$ $$ =i(\bar\Phi_{\alpha
B}\Phi_{\beta A}-\bar\Phi_{\beta B}\Phi_{\alpha A})=
2i\nabla_{[\alpha}(\bar\Phi_B\nabla_{\beta]}\Phi_A).$$ Hence, the
identity~(\ref{twistproperty1}) can be written in the following
form
\begin{equation}
\nabla_{[\alpha}\omega_{\beta]AB}+
2i\nabla_{[\alpha}(\bar\Phi_B\nabla_{\beta]}\Phi_A)=0.\label{twistproperty2}
\end{equation}
From Eq.~(\ref{twistproperty2}) we conclude that there exists a
matrix potential $Y$ such that $$ \nabla_\alpha
Y_{AB}=\omega_{\alpha AB}+ 2i\bar\Phi_B\nabla_\alpha\Phi_A. $$ The
last equation gives
\begin{equation}
e_{\mu\nu\theta\alpha}\nabla^\alpha Y_{AB}K_C^\theta K_D^\nu=
e_{\mu\nu\theta\alpha}\omega^\alpha_{AB}K_C^\theta K_D^\nu+
2i\bar\Phi_B e_{\mu\nu\theta\alpha}\nabla^\alpha\Phi_A
K_C^\theta K_D^\nu.\label{twistpotential1}
\end{equation}

It can be shown that the condition~(\ref{cyk}) is equivalent to
the existence of a two-surface which is orthogonal to both Killing
fields. From now on we chose a coordinate system adopted to the
Killing fields, $K_0=\frac{\partial}{\partial x_0},
\;K_1=\frac{\partial}{\partial x_1},\;x_0=t,\;x_1=\varphi$. Then
\begin{equation}
ds^2=\gamma_{ab}dx^adx^b+g_{AB}dx^Adx^B,
\end{equation}
where $\gamma_{ab}$ and $g_{AB}$ don't depend on $t, \phi$. From
the condition~(\ref{cyk}) we derive that $F^\dagger$ can be
represented in the form
\begin{equation}
F_{\alpha\beta}^\dagger=2 g^{AB}\Phi_{A[\alpha}K_{\beta]B}.
\end{equation}
Here $g^{AB}$ denotes $(g^{-1})_{AB}$. The self-dual property of
$F^\dagger$ leads to the identity
\begin{equation}
g^{AB}e_{\alpha\beta\mu\nu}\Phi^\mu_A K^\nu_B K^\beta_C=i\Phi_{\alpha C}.
\label{selfdual}
\end{equation}

Restricting the equation~(\ref{twistpotential1}) to the
two-surface that is orthogonal $K_A$ and using
Eq.~(\ref{omegaproj}) we obtain
\begin{equation}
\rho\epsilon_{CD}\ast dY_{AB}=
g_{AD}dg_{BC}-g_{AC}dg_{BD}+2i\rho\epsilon_{CD}\bar\Phi_B\ast d\Phi_A.
\label{einsteinreduced}
\end{equation}
Here $\ast$ is the Hodge operator with respect to $\gamma_{ab}$
and $$ -\rho^2=\det
g,\;\;\epsilon_{01}=1,\;\;\epsilon_{AB}=-\epsilon_{BA}. $$
Projection of the identity~(\ref{selfdual}) to the same surface
gives
\begin{equation}
\rho g^{AB}\epsilon_{BC}\ast d\Phi_A=id\Phi_C.\label{maxwellreduced}
\end{equation}
In the matrix notations, Eqs.~(\ref{einsteinreduced}) and
(\ref{maxwellreduced}) can be written as
\begin{equation}
\rho\ast dY=-g\epsilon dg+2i\rho\ast d\Phi\Phi^\ast \label{eqonY}
\end{equation}
and
\begin{equation}
d\Phi=i\rho\epsilon g^{-1}\ast d\Phi.
\label{eqonPhi}
\end{equation}
Here $\Phi$ is a column vector and  $\Phi^\ast$ is its Hermitian
conjugation; $\Phi^\ast=(\bar\Phi_0, \bar\Phi_1)$.

With the help of Eq.~(\ref{eqonY}) and Eq.~(\ref{eqonPhi}) we can
show that the Einstein-Maxwell equations take the form
\cite{alexeev}
\begin{equation}\label{eqonG}
d(\rho G^{-1}\ast dG)=0,
\end{equation}
where
\begin{equation}\label{defofG}
G=\left(\begin{array}{ll} g+2\Phi\Phi^{*} & \Phi\\ \Phi^{*} & 1/2
\end{array}\right);\;\;\;
G^{-1}=\left( \begin{array}{ll} g^{-1} & -2g^{-1}\Phi \\
-2\Phi^{*}g^{-1} & 4\Phi^{*}g^{-1}\Phi+2 \\ \end{array} \right).
\end{equation}
Notice that
\begin{equation}
G^\ast=G,\;\;\;\det G=-\frac{\rho^2}{2}.
\end{equation}
To prove (\ref{eqonG}), we note first that
\begin{equation}
\rho G^{-1}\ast dG=
\left(\begin{array}{ll} \epsilon dY & i\epsilon d\Phi\\
-2\Phi^{*}\epsilon dH+2id(\Phi^{*}\epsilon g) & -2i\Phi^{*}\epsilon d\Phi
\end{array}\right)
\end{equation}
where $H$ is Kinnersley's potential \cite{kin77}, $dH=dY+idg$. So,
we need only to check the validity of the following identities
$$d\Phi^{*}\wedge\epsilon d\Phi=d\Phi^{*}\wedge\epsilon dH=0.$$
From Eq.~(\ref{eqonY}) and Eq.~(\ref{eqonPhi}), it is easy to see
that
\begin{equation} dH=i\rho\epsilon g^{-1}\ast
dH.\label{identityonH}
\end{equation}
Then, we have $$d\Phi^{*}\wedge\epsilon d\Phi=\ast
d\Phi^{*}\wedge\epsilon\ast d\Phi= -d\Phi^{*}\wedge\epsilon
d\Phi=0$$ and $$d\Phi^{*}\wedge\epsilon dH=\ast
d\Phi^{*}\wedge\epsilon\ast dH=- d\Phi^{*}\wedge\epsilon dH=0.$$
Here we use the standard property of the Hodge operator with
respect to $\gamma_{ab}$ and the
equations~(\ref{eqonPhi}),~(\ref{identityonH}).

It is worth mentioning that the same equation~(\ref{eqonG}) was
derived in Ref.~\cite{xan} but the matrices $G$ defined there and
here are different.

We now turn to discussion of the boundary conditions. From
Eq.~(\ref{eqonG}) one sees that $d\ast d\rho=0$. Let $\rho$ and
$z$, where $z$ defined by $dz=\ast d\rho$, be a coordinate system
of the two-surface. Then
$$\gamma_{ab}dx^adx^b=f(z,\rho)(d\rho^2+dz^2),\;\;\;dz=\ast
d\rho,\;\;\; d\rho=-\ast dz.$$ One can prove that the function
$f(z,\rho)$ is uniquely determined by the matrix $G$. We won't
write these equations since they aren't need us in this paper.
Virtually, the factor $f$ is irrelevant to the problem of N black
holes since we can compute the interaction force using the
properties of the metric coefficient $g_{11}$
Ref.~\cite{varzugin98}. Let us remark that the Hodge operator is
conformally invariant and independent of $f$.

Define $$g_{00}=-V,\;\;\;g_{11}=X,\;\;\; g_{01}=W.$$ The event
horizon and the axis of symmetry can be described as follows. The
set of points with $\rho=0$ and $X=0$ is the symmetry axis while
the set of points with $\rho=0$ and $X>0$ is the event horizon.
Denote through $z_k$ ($k=1,\dots, 2N$ N-black holes)
$z$-coordinates of the intersection points of the horizon and the
symmetry axis. Then an interval $I_i=[z_{2i}, z_{2i-1}]$
corresponds to the horizon of $i$th black hole. The symmetry axis
components we denote by $\Gamma_i$ for $i=1,\dots, N+1$ where
$\Gamma_i=(z_{2i-1},z_{2i-2})$. It is assumed that $z_0=-\infty$
and $z_{2N+1}=+\infty$.

We pass to a new coordinate system in a neighborhood of $i$th
black hole, $$\rho^2=(\lambda^2-m_i^2)(1-\mu^2),\;\;\;
m_i={z_{2i}-z_{2i-1}\over2},$$
$$z-{z_{2i}+z_{2i-1}\over2}=\lambda\mu.$$ For this coordinate
system the horizon and the symmetry axis regularity conditions can
be written in the form \cite{carter}
\begin{equation}
\label{bconong}
\pmatrix{1&\Omega_i\cr0&1\cr}g\pmatrix{1&0\cr\Omega_i&1\cr}=
\pmatrix{(\lambda^2-m_i^2)\hat V(\lambda,\mu)&\rho^2\hat W(\lambda,\mu)\cr
\rho^2\hat W(\lambda,\mu)&
(1-\mu^2)\hat X(\lambda,\mu)\cr},
\end{equation}
\begin{equation}
\label{bcononphi}
(\Phi_0+\Omega_i\Phi_1)=\Phi^H_i+(\lambda^2-m_i^2)\hat\Phi_0(\lambda,\mu),
\;\;\;\Phi_1=\Phi^A_i+(1-\mu^2)\hat\Phi_1(\lambda,\mu),
\end{equation}
where $\hat\Phi_0, \hat\Phi_1, \hat X, \hat V$ ¨ $\hat W$ are
smooth functions nowhere equal to zero, and $\Omega_i, \Phi^H_i$ ¨
$\Phi^A_i$ are some constants. The second condition of
(\ref{bcononphi}) is understood as the regularity condition for
the electromagnetic field in the vicinity of the symmetry axis
components, $\Phi_1=\Phi_i^A$ in $i$th component $(i=1,\dots,
N+1)$ of the symmetry axis. The constant $\Omega_i$ is the angular
velocity of $i$th black hole. Note that one can derive the
conditions for the imaginary part of $\Phi$ from the results of
Ref.~\cite{carter} using the equation~(\ref{eqonPhi}).

In addition, we require that
\begin{equation}
\label{goninfty}
X=\rho^2(1+O(1/r)),\;\;\;W=\rho^2O(1/r^3),\;\;\;V=1+O(1/r),
\end{equation}
\begin{equation}
\label{phioninfty}
\Phi_0=O(1/r),\;\;\;\Phi_1=O(1),
\end{equation}
as $r\to\infty$ where $r=\sqrt{z^2+\rho^2}$ and the
asymptotics~(\ref{goninfty}) and~(\ref{phioninfty}) are
differentiable. In particular, ${\Phi_1}_{,z}=O(1/r)$ and
${\Phi_1}_{,\rho}=O(1/r)$. The function $\Phi_1$ is given up to an
additive constant which we fix requiring that
\begin{equation}
\label{norphi}
\lim_{z\to-\infty}\Phi_1(z,\rho)=\Phi_1^A=i\bar q,\;\;\;
\lim_{z\to+\infty}\Phi_1(z,\rho)=\Phi_{N+1}^A=-i\bar q,
\end{equation}
where $q$ is the total charge of the system (see below). The last
condition is compatible with the equations~(\ref{eqonG}).

The charge of $j$th black hole is defined by an integral
\begin{equation} \bar
q_j=Q_j-iP_j=-\frac{1}{4\pi}\int_{S_j}F^{\dagger\mu\nu}dS_{\mu\nu},
\end{equation}
where $Q_j$ is the electric charge, $P_j$ is the magnet monopole
charge and $S_j$ is a two-surface surrounding $j$th black hole. We
assume that the two-surface belongs to a spacelike hypersurface
which is invariant under the action of the rotational isometry and
its normal orientation is chosen in the direction of the space
infinity. Let $S_j$ be a surface of revolution of a curve $C_j$.
Then
\begin{equation}
\bar q_j=\frac{i}{2}\int_{C_j}d\Phi_1=\frac{i}{2}(\Phi_{j+1}^A-\Phi_j^A).
\end{equation}
Taking this into account we see that the constant $q$ in the
normalization condition~(\ref{norphi}) is the sum of the charges
of all black holes $q=\sum q_i$.

The mass and the angular momentum of the system are defined by
integrals
\begin{equation}
M=-\frac{1}{4\pi}\int_{S_\infty}k^{\mu;\nu}dS_{\mu\nu},\;\;\;
L=\frac{1}{8\pi}\int_{S_\infty}m^{\mu;\nu}dS_{\mu\nu},
\end{equation}
where $S_\infty$ is the space infinity which is a two-sphere.
Using Stokes' theorem and the Einstein-Maxwell equations we obtain
\begin{equation}
L=\sum_iL_i,\;\;\;L_i=\frac{1}{8\pi}\int_{H_i}{\cal L}^{\mu\nu}dS_{\mu\nu},
\;\;\;{\cal L}_{\mu\nu}=m_{\mu;\nu}-\Phi_1\bar F^\dagger_{\mu\nu},
\end{equation}
and
\begin{equation}
M=\sum_iM_i,\;\;\;M_i=-\frac{1}{4\pi}\int_{H_i}{\cal M}^{\mu\nu}dS_{\mu\nu},
\;\;\;{\cal M}_{\mu\nu}=k_{\mu;\nu}-\Phi_0\bar F^\dagger_{\mu\nu}.
\end{equation}
Here $H_i$ is the horizon component. Since the horizon is a
two-surface of revolution we get
\begin{equation}
L_i=-\frac{1}{8}(\bar Y_{11}(z_{2i})-\bar Y_{11}(z_{2i-1})),\;\;\;
M_i=\frac{1}{4}(\bar Y_{10}(z_{2i})-\bar Y_{10}(z_{2i-1})).
\end{equation}
A simple corollary of the boundary conditions~(\ref{bconong})
and~(\ref{bcononphi}) is the identity $Y_{10,z}+\Omega_j
Y_{11,z}|_{I_j}=2+2i\bar\Phi^H_j\Phi_{1,z}$. Using it we find that
\begin{equation}
\label{law1}
M_i=m_i+2\Omega_iL_i+\Phi^H_iq_i.
\end{equation}
Recall $m_i=(z_{2i}-z_{2i-1})/2$.

Under the gauge transformation $\Phi_1\to\Phi_1+a$ the quantities
$M_i$ and $L_i$ change as $M_i\to M_i-a\Omega_iq_i$ and $L_i\to
L_i-\frac{1}{2}aq_i$, respectively. We define the physical mass
and the angular momentum of one black hole by formulas
\begin{equation}
M_i^{\rm ph}=M_i+\Omega_iq_i\frac{\Phi_{i+1}^A+\Phi_{i}^A}{2},\;\;\;
L_i^{\rm ph}=L_i+q_i\frac{\Phi_{i+1}^A+\Phi_{i}^A}{4},
\end{equation}
$$\frac{\Phi_{i+1}^A+\Phi_i^A}{2}=i\sum_{i+1}^N\bar
q_k-i\sum_1^{i-1}\bar q_k.$$ Notice that $M_i^{\rm ph}$ and
$L_i^{\rm ph}$ are real and exactly equal to the mass and the
angular momentum of the isolated black hole, respectively. In
addition, $M_i^{\rm ph}$ and $L_i^{\rm ph}$ satisfy
Eq.~(\ref{law1}) hence the product $q_i\Phi_i^H$ is a real
quantity.

In the end of this section we state the boundary conditions for
Eqs.~(\ref{eqonG}). From Eqs.~(\ref{bconong}), (\ref{bcononphi})
we deduce that
\begin{equation}
\label{cononA}
\rho G_{,\rho}G^{-1}=\left(\begin{array}{lll}
0& -\bar Y_{00,z}& 2\Phi_i^A\bar Y_{00,z}\\
0& -\bar Y_{01,z}& 2\Phi_i^A\bar Y_{01,z}\\
0 &  i\bar\Phi_{0,z}& -2i\Phi_i^A\bar\Phi_{0,z}\\
\end{array}\right),
\end{equation}
as $\rho=0$ and $z\in\Gamma_i$ while
\begin{equation}
\label{cononH}
\hat\Omega_i\rho G_{,\rho}G^{-1}\hat\Omega_i^{-1}=\left(\begin{array}{lll}
\bar Y_{10,z}& 0& -2\Phi_i^H\bar Y_{10,z}\\
\bar Y_{11,z}& 0& -2\Phi_i^H\bar Y_{11,z}\\
-i\bar\Phi_{1,z} & 0& 2i\Phi_i^H\bar\Phi_{1,z}\\
\end{array}\right),\;
\hat\Omega_i=\left(\begin{array}{ccc}
1&\Omega_i&0\\
0&1&0\\
0&0&1\\
\end{array}\right),
\end{equation}
as $\rho=0$ and $z\in I_i$ and $\rho G_{,z}G^{-1}=O(1)$ as
$\rho\to0$.

\section{Constraint equations}

Not all of the parameters introduced in the previous section are
independent. In this section we discuss a way of deriving the
relations between the masses, the angular momenta, the angular
velocities, the charges, the distance parameters, the values of
the electromagnetic field potential at the horizon and at the
symmetry axis.

The system of equations~(\ref{eqonG}) is a compatibility condition
of the following pair of matrix linear differential equations
\cite{belzhah78}
\begin{equation}
\label{para}
D_1\psi={\rho^2G_{,z}G^{-1}-\omega\rho G_{,\rho}G^{-1}\over
\omega^2+\rho^2}\psi,\;\;
D_2\psi={\rho^2G_{,\rho}G^{-1}+\omega\rho G_{,z}G^{-1}\over\omega^2+
\rho^2}\psi,
\end{equation}
where $D_1, D_2$ are commuting differential operators,
$$D_1=\partial_z-{2\omega^2\over\omega^2+\rho^2}\partial_\omega,
\;\;D_2=\partial_\rho+{2\omega\rho\over\omega^2+\rho^2}
\partial_\omega,$$
and $\omega$ is a complex parameter that doesn't depend on the
coordinates. Studying the non-charged black holes in
Ref.~\cite{varzugin97} we have showed that well-known methods of
investigation of completely integrable equations \cite{TF} can be
applied to the problem of N black holes. Moreover, this problem
occurs to be explicitly solvable provided that the relations
between the parameters are found. We proceed discussing the
principle steps of Ref.~\cite{varzugin97}.

Let $\omega$ be a root of the equation
\begin{equation}
\label{omega}
\omega^2-2\omega(k-z)-\rho^2=0,
\end{equation}
where, now, $k$ is an independent spectral parameter. From
Eq.~(\ref{omega}) it follows that
$$\partial_z\omega=-{2\omega^2\over\omega^2+\rho^2},\;\;
\partial_\rho\omega={2\omega\rho\over\omega^2+\rho^2}.$$
Hence, $\psi^\prime(k)=\psi(\omega(k))$ is a solution of the
linear equation
\begin{equation}
\label{paraz}
\partial_z\psi^\prime(k)={\rho^2G_{,z}G^{-1}-\omega\rho G_{,\rho}G^{-1}\over
\omega^2+\rho^2}\psi^\prime(k).
\end{equation}
Invariance of Eq.~(\ref{omega}) with respect to the transformation
$\omega \rightarrow -\rho^2/\omega$ allows us to fix a branch of
the function $\omega(k)$ through the inequality $|\omega|>\rho$.
Then, from Eq.~(\ref{omega}) we see that$$
\omega\rightarrow2(k-z),\;\rho\rightarrow0,\;\;\;
\omega\rightarrow2(k-z),\; z\rightarrow\infty.\; $$

After choosing a branch of $\omega$, we can define the monodromy
matrix $T(z,y)$, which, by definition, is a solution to
Eq.~(\ref{paraz}) such that $T(y,y)=I.$  Using the boundary
conditions (\ref{cononA}) and (\ref{cononH}), $T(z,y)$ can be
explicitly evaluated at $\rho=0$. More precisely, for $\rho=0$ and
$z,y\in\Gamma_i$ one obtains $$ T(z,y)=\left(\begin{array}{ccc}
1&\frac{\bar Y_{00}(z)-\bar Y_{00}(y)}{2(k-y)}&-\Phi_i^A
\frac{\bar Y_{00}(z)-\bar Y_{00}(y)}{k-y}\\
0&\frac{k-z}{k-y}-i\Phi_i^A\frac{\bar\Phi_0(z)-\bar\Phi_0(y)}{k-y}&
2\Phi_i^A(\frac{z-y}{k-y}+i\Phi_i^A\frac{\bar\Phi_0(z)-\bar\Phi_0(y)}{k-y})\\
0&-i\frac{\bar\Phi_0(z)-\bar\Phi_0(y)}{2(k-y)}&
1+i\Phi_i^A\frac{\bar\Phi_0(z)-\bar\Phi_0(y)}{k-y}\\
\end{array}\right),
$$ while for $\rho=0$ and $z,y\in I_i$ one gets $$
T(z,y)=\hat\Omega_i^{-1}\left(\begin{array}{ccc}
\frac{k-z}{k-y}+i\Phi_i^H\frac{\bar\Phi_1(z)-\bar\Phi_1(y)}{k-y}&
0&
2\Phi_i^H(\frac{z-y}{k-y}-i\Phi_i^H\frac{\bar\Phi_1(z)-\bar\Phi_1(y)}{k-y})\\
-\frac{\bar Y_{11}(z)-\bar Y_{11}(y)}{2(k-y)}& 1&
\Phi_i^H\frac{\bar Y_{11}(z)-\bar Y_{11}(y)}{k-y}\\
i\frac{\bar\Phi_1(z)-\bar\Phi_1(y)}{2(k-y)}& 0&
1-i\Phi_i^H\frac{\bar\Phi_1(z)-\bar\Phi_1(y)}{k-y}\\
\end{array}\right)\hat\Omega_i
$$

The reduced monodromy matrix $T(k)$ is defined by
\begin{equation}
\label{T}
T(k)=\lim_{y\rightarrow -\infty,z\rightarrow +\infty}
e_+^{-1}(k,z)T(z,y)e_-(k,y),
\end{equation}
where $$ e_\pm(k,z)=\left(\begin{array}{ccc}
1&0&0\\0&\omega(k,z)&\mp2i\bar q\\0&0&1\\
\end{array}\right).
$$ The matrix $T(k)$ doesn't depend on $\rho$. It is a consequence
of the conditions~(\ref{goninfty}), (\ref{phioninfty}) and
(\ref{norphi}). Furthermore, as $z\in\Gamma_{N+1}$ and
$y\in\Gamma_1$ the monodromy matrix $T(z,y)$ can be presented in
the form $$T(z,y)=T(z,z_{2N})T(z_{2N},z_{2N-1})\dots T(z_1,y).$$
Taking the last formula and the known expressions for $T(z,y)$ at
$\rho=0$ into account one finds the limit~(\ref{T}) easily. We
write the result in the following form
\begin{equation}
T(k)=K_{N+1}T_NK_{N}T_{N-1}\dots K_2T_1K_1.
\end{equation}
Here $$ T_j=\left(\begin{array}{ccc}
1-\frac{2M_j}{k-z_{2j-1}}&-4\Omega_jM_j&\frac{4M_j\phi_{2j-1}}{k-z_{2j-1}}\\
\frac{2L^{\rm ph}_j-i|q_j|^2}{(k-z_{2j})(k-z_{2j-1})}&
\frac{k-z_{2j-1}}{k-z_{2j}}+\frac{2(2L_j^{\rm
ph}-i|q_j|^2)\Omega_j}{k-z_{2j}}& \frac{2i\bar q_j}{k-z_{2j}}-
\frac{2(2L_j^{\rm
ph}-i|q_j|^2)\phi_{2j-1}}{(k-z_{2j})(k-z_{2j-1})}\\
-\frac{q_j}{k-z_{2j-1}}&-2q_j\Omega_j&1+\frac{2q_j\phi_{2j-1}}{k-z_{2j-1}}\\
\end{array}\right)
$$ and $$ K_j=\left(\begin{array}{ccc}
1&D_j&0\\0&1&0\\0&-i(\bar\phi_{2j-1}-\bar\phi_{2j-2})&1\\
\end{array}\right),
$$ where $D_j=\bar Y_{00}(z_{2j-1})-\bar Y_{00}(z_{2j-2})$ and
$\phi_k=\Phi_0(z_k)$. Recall $z_0=-\infty$ and $z_{2N+1}=+\infty$.
We note also that $$\phi_{2j}=\Phi^H_j-\Omega_j\Phi_{j+1}^A,\;\;\;
\phi_{2j-1}=\Phi^H_j-\Omega_j\Phi_j^A$$ and
$2L_j+\Phi^A_{j+1}q_j=2L_j^{\rm ph}-i|q_j|^2$.

The equations~(\ref{para}) are invariant with respect to the
transformations $$\psi(\omega)\rightarrow
G(\psi^\ast(-\frac{\rho^2}{\bar\omega}))^{-1},\;\;\;
\psi(\omega)\rightarrow
\Phi_g(\omega)\psi^\ast(\bar\omega)^{-1},$$ where $$
\Phi_g(\omega)=\left(\begin{array}{ccc}
2|\Phi_0|^2&2\Phi_0\bar\Phi_1-i\omega&\Phi_0\\
2\Phi_1\bar\Phi_0+i\omega&2|\Phi_1|^2&\Phi_1\\
\bar\Phi_0&\bar\Phi_1&1/2\\
\end{array}\right).
$$ Thus, the reduced monodromy matrix $T(k)$ has to satisfy
\begin{equation}
\label{Tsym}
T(k)=\left(\begin{array}{cc}-I&0\\0&\frac{1}{2}\\\end{array}\right)
T^\ast(\bar k)
\left(\begin{array}{cc}-I&0\\0&2\\\end{array}\right)
\end{equation}
and
\begin{equation}
\label{Tsym2}
T(k)=\left(\begin{array}{cc}\sigma_y&0\\0&\frac{1}{2}\\\end{array}\right)
T^\ast(\bar k)^{-1}
\left(\begin{array}{cc}\sigma_y&0\\0&2\\\end{array}\right).
\end{equation}
Here $I$ is the unit 2x2 matrix, $\sigma_y$ is the Pauli matrix
and $T^\ast$ is the Hermitian conjugation of $T$. The matrix
$T(k)$ can not satisfy Eq.~(\ref{Tsym}) if the physical parameters
are arbitrary. This equation~(\ref{Tsym}) defines the relations
between the parameters. Unfortunately, these relations are quite
complicated and we can study them in particular cases only.

Let us consider one black hole first. From Eq.~(\ref{Tsym}) we
obtain $$D_1=2\Omega_1M_1-\frac{i|q_1|^2M_1\Omega_1}{L_1},\;\;\;
D_2=2\Omega_1M_1+\frac{i|q_1|^2M_1\Omega_1}{L_1}$$ and
$$\phi_1=(\phi^H-i\Omega_1)\bar
q_1,\;\;\;\phi_2=(\phi^H+i\Omega_1)\bar q_1,
\;\;\;\phi^H=\frac{L_1+M_1\Omega_1|q_1|^2}{2L_1M_1}.$$ Then, the
equations~(\ref{Tsym}) and (\ref{law1}) are eventually reduced to
the familiar formulas for the mass and the angular velocity of one
black hole, viz
$$\Omega_1=\frac{a}{(M_1+m_1)^2+a^2},\;\;M_1^2=m_1^2+a^2+|q_1|^2,
\;\;a=\frac{L_1}{M_1}.$$

We can solve Eqs.~(\ref{Tsym}) for the case of two non-rotating
black holes($\Omega_1=\Omega_2=0$) with the charges of the magnet
monopoles equal to zero $(P_1=P_2=0)$ as well. In this case,
$L_1^{\rm ph}= L_2^{\rm ph}=0$, $M_1^{\rm ph}=M_1$ and $M_2^{\rm
ph}=M_2$, while Eqs.~(\ref{Tsym}) and (\ref{law1}) are eventually
reduced to a system of two equations
\begin{equation} \label{massy}
M_1=m_1+\phi^H_1Q_1^2,\;\;\;M_2=m_2+\phi^H_2Q_2^2,
\end{equation}
where $$\phi_1^H=\frac{1}{M_1+m_1}
\left(1+\frac{2M_1Q_2-2M_2Q_1}{Q_1(M_1+M_2+R)}\right), $$ $$
\phi_2^H=\frac{1}{M_2+m_2}
\left(1-\frac{2M_1Q_2-2M_2Q_1}{Q_2(M_1+M_2+R)}\right), $$ and $R$
is a distance parameter chosen such a way that $z_1=-m_1, z_2=m_1,
z_3=R-m_2$ and $z_4=R+m_2$. For completeness, we note that
$$D_1=-i{\Phi_1^H}^2,\;\;\;D_2=-i({\Phi_2^H}^2-{\Phi_1^H}^2),
\;\;\;D_3=i{\Phi_2^H}^2$$ and
$$\Phi_1^H=\phi_1^HQ_1,\;\;\;\Phi_2^H=\phi_2^HQ_2.$$

The system of Eqs.~(\ref{massy}) is one of the main results of
this paper. It gives the relations between parameters $M_1, M_2$
and $m_1, m_2$. It is preferred to consider $m_i$, which are the
irreducible masses, as independent parameters of the problem.
First, because they describe the event horizon. Secondly, in the
general relativity, there is no notion of the energy density
hence, we think, there is no sense to fix the mass of $i$th black
hole. Moreover, the total mass $M= M_1+ M_2$ contains the energy
of the electromagnetic field surrounding the black holes as well.
Let $m_i$ be independent parameters then in the general case the
masses $M_i^{\rm ph}$ become functions of the distance parameter.
These functions must be derived from Eqs.~(\ref{Tsym}) and
(\ref{law1}).

In the particular case of Eqs.~(\ref{massy}) we see that $M_1$ and
$M_2$ don't depend on $R$ only if $M_1Q_2=M_2Q_1$. This is true if
$m_1Q_2=m_2Q_1$. The quantities $Q_i/2m_i$ have the sense of
charge per unit length. Hence, if the charge density of two black
holes treated as one dimensional objects, namely, the intervals
$I_i$ are different the total mass of the system depends on the
distance between the black holes.

\section*{Acknowledgements}
This work was partly supported by RFBR grant No. 00-01-00480.

\end{document}